\documentclass[12pt]{article}

\newlength{\uppermar}
\newlength{\lowermar}
\newlength{\leftmar}
\newlength{\rightmar}

\usepackage[centertags]{amsmath}
\usepackage{amsfonts}
\usepackage{amssymb}
\usepackage{amsthm}
\usepackage[rflt]{floatflt}
\usepackage{feynmf,multibox}
\usepackage[russian,english]{babel}
\usepackage{calc}

\DeclareMathOperator{\cycle}{cycle}

\def\gh{\mathrm{gh}}
\def\se{\stackrel{\ast}{\eta}{}\mskip-5mu}
\def\sx{\stackrel{\ast}{x}{}\mskip-5mu}
\def\sc{\stackrel{\ast}{c}{}\mskip-5mu}

\begin{document}

\setlength{\unitlength}{1mm}


\title{BRST theory without Hamiltonian and Lagrangian}

\author
{ S.L. Lyakhovich\thanks{sll@phys.tsu.ru} and A.A.
Sharapov\thanks{sharapov@phys.tsu.ru.}\\ \textit{\small Department
of Theoretical Physics, Tomsk State University, Tomsk 634050,
Russia.}}

\maketitle

\begin{abstract}
We consider a generic gauge system, whose physical degrees of
freedom are obtained by restriction on a constraint surface
followed by factorization with respect to the action of gauge
transformations; in  so doing, no Hamiltonian structure or action
principle is supposed to exist. For such a generic gauge system we
construct a consistent BRST formulation, which includes the
conventional BV Lagrangian and BFV Hamiltonian schemes as
particular cases. If the original manifold carries a weak Poisson
structure (a bivector field giving rise to a Poisson bracket on
the space of physical observables) the generic gauge system is
shown to admit deformation quantization by means of the Kontsevich
formality theorem. A sigma-model interpretation of this
quantization algorithm is briefly discussed.
\end{abstract}

\section{Introduction}

The standard BRST theory of gauge systems \cite{ht} starts from
either Lagrangian formulation of the original dynamics (the BV
method \cite{bv}) or Hamiltonian constrained formalism (the BFV
method \cite{bfv}). This paper is about constructing a BRST
embedding for the gauge systems, whose physical degrees of freedom
are defined merely by restriction to a smooth submanifold (called
shell) accompanied  by factorization with respect to the action of
an on-shell integrable vector distribution (gauge algebra
generators). In this quite general setting we start with, the
shell can be thought of, for example, as defined either by
equations of motion (not necessarily Lagrangian) or by a
constraint surface (not necessarily Hamiltonian). Accordingly, the
gauge algebra generators, foliating the shell into the gauge
orbits, are required neither to annihilate some action functional
nor to be induced by a set of Hamiltonian first-class constraints.

In most cases of gauge systems the Lagrangians  are known, though
the absence of an action principle is not uncommon: the equations
for interacting higher-spin gauge fields \cite{hs}, for example,
do not admit the action principle, at least in their present form.
In the recent paper \cite{BGST}, it is remarked that any given
BRST differential in the space of trajectories allows to identify
equations of motion and gauge symmetries independently of the
question of the existence of a Lagrangian for these equations.
Using this observation, a general BRST formulation for Vasiliev's
method of unfolding \cite{hs} was given in \cite{BGST} for the
case of free field theories associated to first-quantized
constrained systems.

In the present paper, proceeding from the understanding that the
physical observables are the functions defined modulo on-shell
vanishing terms and remained constant along the gauge orbits, we
formulate a generic gauge theory with no reference to any
Lagrangian or Hamiltonian structure. As a next step we suggest a
uniform scheme for the BRST embedding of the generic gauge system
whose dynamics is defined just by equations of motion and
constraints, with no Poisson bracket or action principle involved.
This more general view may also have some impact on understanding
of the gauge systems which can admit a conventional Lagrangian or
Hamiltonian description as well. In particular, in such a general
setting, the gauge generators can be much less tightly bound to
the shell, than it is usually required in constrained Hamiltonian
dynamics. Being applied to Hamiltonian systems, this formalism
allows for factorization of mixed-class constrained surfaces by
any on-shell integrable transformations, not only by the
Hamiltonian action of the first class constraints, as it is
usually implied in the conventional BRST-BFV scheme. Some
approaches of this sort to the mixed class systems have been
discussed, for instance, in Ref. \cite{blm}.

In order to quantize the generic gauge system, we study a
possibility to endow only the space of physical observables with a
Poisson algebra structure; in so doing, no Poisson structure is
supposed to exist on the original space of the system or on its
BRST extension. The existence of the Poisson brackets on the
entire space and the off-shell compatibility of the Poisson
structure with the gauge algebra seem to be excessive requirements
from the viewpoint of a consistent physical interpretation and
deformation quantization of gauge systems.

Geometrically, the usual requirements of BFV method for the
Poisson brackets to possess the off-shell Jacobi identity and the
gauge generators to be off-shell Hamiltonian vector fields seem as
superficial as demanding the gauge generators to form an off-shell
integrable distribution. The later requirement has been recognized
as an excessive restriction long ago. In Refs. \cite{bt},
\cite{bm} weak Poisson brackets, satisfying the Jacobi identity
only on shell, have been already studied in the BRST framework.
The relevance of the weak Poisson brackets to the covariant
quantization of superstrings was discussed \cite{dgl}. In this
work we find that the concept of the weak Poisson brackets can be
further relaxed for the gauge systems: it is sufficient to have
the on-shell Jacobi identity only for physical observables (i.e.
gauge invariant values), not for arbitrary functions.

We show that such a generalized notion of the weak Poisson
structure fits well to the general concept of gauge systems
proposed in Ref. \cite{LS}. More precisely, it is possible to
generate all the structure relations underlying both the gauge
symmetry and the weak Poisson structure from a single master
equation $(S,S)=0$, where $S$ is an even ghost-number-2 function
on an appropriately chosen anti-Poisson manifold. The physical
dynamics is specified by means of an odd, ghost-number-1 function
$H$ subject to the equation $(S,H)=0$. Given the generic gauge
system on the original manifold $M$, the functions $S$ and $H$ are
systematically  constructed by solving the master equations on the
properly ghost extended $M$.

The functions $S$ and $H$ serve as prerequisites to perform the
deformation quantization of the generic gauge system. The quantum
counterpart of the weak Poisson structure is a weakly associative
$\ast$-product. The latter can be constructed by applying the
Kontsevich formality theorem. The ``weakness'' means that upon the
quantum reduction the $\ast$-product induces an associative
quantum multiplication for physical observables. It turns out that
the direct application of the formality map to $S$ and $H$ may
face the problem of quantum anomalies (i.e. the values which break
the existence of quantum reduction) and we comment on how to
remove these anomalies by a proper redefinition of $S$ and $H$.

Finally, we show that, as for usual Poisson manifolds, the
deformation quantization of the weak Poisson structures can be
explored in terms of a 2-dimensional topological sigma-model
associated to $S$. This approach can be used as the basis for
definition and (non-rigor) quantization of the generic gauge
systems associated with non-canonical antibrackets
$(\,\cdot\,\,,\cdot\,)$. The non-canonical antibrackets become
inevitable, e.g., when the normal bundle to the shell and/or the
tangent bundle of the on-shell gauge foliation admit no flat
connection.

\section{Generic constraints and gauge algebra}

In any gauge theory, true physical degrees of freedom appear as a
result of reduction of the original configuration or phase-space
manifold. Two different ways are known for the reduction: i) a
direct restriction of the dynamics to a submanifold (which we call
a shell), and ii) factorization with respect to the action of
vector fields generating the gauge invariance on shell. The shell
can be understood, for example, as a constraint surface in
constrained Hamiltonian dynamics or as a mass-shell of Lagrangian
gauge theory. The compatibility conditions between restriction and
factorization requires the gauge generators to define an on-shell
integrable distribution. In other words, the generic gauge
transformations are not necessarily integrable in the entire
configuration/phase space, foliating only the shell into the gauge
orbits. The physical observables are defined modulo on-shell
vanishing terms and are required to be constant along the gauge
orbits. In this section, we describe the algebra of the physical
observables of the generic gauge system where the physical
dynamics emerges from the reduction of base manifold to any shell
(not necessarily Hamiltonian constraint surface or Lagrangian
mass-shell) followed by factorization with respect to any on-shell
integrable distribution.

Let $M$ be a smooth manifold with local coordinates $x^i$.
Consider a nonempty subset $\Sigma\subset M$, called the shell,
which is determined by a system of $m$ equations
\begin{equation}\label{T}
    T_a (x) =0 \, , \quad a=1,\ldots , m,
\end{equation}
called the constraints. The standard regularity condition
\begin{equation}\label{r1}
    \mathrm{rank}\left.\left(\frac{\partial T_a}{\partial x^i} \right)\right|_\Sigma=m
\end{equation}
ensures that $\Sigma$ is a smooth submanifold of
$\mathrm{codim}\Sigma=m$. In the language of constrained dynamics,
this means that the constraints $T_a(x)$ are supposed to be
irreducible:
\begin{equation}
\label{irr}
   Z^a(x) T_a (x) =0 \, \, \Rightarrow \, \,\exists\, A^{ab}(x)= - A^{ba}(x) : \, Z^a = A^{ab} T_b \,.
\end{equation}
If desired, $\Sigma$ can be regarded as a mass-shell of a
Lagrangian gauge system (or the shell of the equations of motion
of a non-Lagrangian theory), though this requires constraints $T_a
$ to be reducible. In other words, the on-shell nonvanishing
null-vectors $Z^a$ have to exist, defining the Noether identities
between the equations of motion\footnote{The generators of
identities do not necessarily coincide with the generators of
gauge symmetries for a non-Lagrangian gauge system.} $T_a=0$. It
is not a problem to extend the consideration to the reducible case
(this will be done in the next work), but in this paper we confine
ourselves with the irreducible case for the sake of technical
simplicity. So it is slightly easier to think of $T_a$ as
constraints (e.g. the Hamiltonian ones).

Consider a set of $n$ vector fields $R_a=R^i_\alpha\partial_i$ on
$M$, which are tangent to $\Sigma$, i.e.
\begin{equation}\label{R}
[R_\alpha,T_a]\equiv R^i_\alpha \partial_iT_a = E_{\alpha a}^b
T_b\,,
\end{equation}
for some $E_{\alpha b}^c(x)\in C^\infty(M)$. Hereafter the square
brackets $[\,\cdot\, ,\,\cdot\,]$ stand for the Schouten
commutator in the exterior algebra $\Lambda(M)=\bigoplus
\Lambda^k(M)$ of smooth polyvector fields on $M$.
\footnote{Mention that the Schouten commutator is the unique
bracket of polyvector fields which agrees with the Lie bracket of
vector fields and is a graded derivation of degree -1 of the
exterior product.} The vector fields $R_\alpha$ are assumed to be
linearly independent on $\Sigma$ and to form an on-shell
integrable distribution. In other words,
\begin{equation}\label{r2}
    \mathrm{rank}(R_\alpha^i)|_\Sigma=n
\end{equation}
and
\begin{equation}\label{rinv}
\begin{array}{c}
   \frac12 [R_\alpha ,  R_\beta] = U_{\alpha\beta}^\gamma R_\gamma +
    T_a A^a_{\alpha\beta} \, ,
    \end{array}
\end{equation}
for some vector fields $A^a_{\alpha\beta} \in \Lambda^1(M)$. The
vector fields $R_\alpha$ will be called gauge generators if the
conditions (\ref{R},\ref{rinv}) are satisfied. The irreducibility
condition (\ref{r2}) can be relaxed, but in this paper we consider
only this simplest case.

Usually the gauge generators are assumed to be either Hamiltonian
vector fields for the first class constraints or the annihilators
of the action functional. In these standard cases, the relations
(\ref{R},\ref{rinv}) hold true and, as is argued below, it is the
condition which defines the gauge system even when no Hamiltonian
or Lagrangian structure is admissible on the original manifold.
Since, in general, the integrability of the vector distribution
$\{R_\alpha\}$ takes place only on the submanifold $\Sigma$,
determined by the system of constraints $T_a$, we will refer to
the involution relations (\ref{R},\ref{rinv}) as a
\textit{constrained gauge algebra}. On shell, the constrained
gauge algebra is nothing but the Lie algebroid, off shell it is
not, in general.

Let $\mathcal{F}(\Sigma)$ denote the foliation associated to the
integrable distribution $\{\bar R_\alpha=R_\alpha|_\Sigma\}$ and
let $N$ be the corresponding  space of leaves (the points of $N$
are just leaves of the foliation $\mathcal{F}(\Sigma)$). In
general, $N$ may not be  a smooth, Hausdorff manifold.
Nonetheless, we can define the space $\mathcal{T}(N)$ of
``smooth'' tensor fields on $N$ as the space of invariant tensor
fields on $\Sigma$:
\begin{equation}\label{}
    L_{\bar R_\alpha} S =0\;\Leftrightarrow \;S\in
    \mathcal{T}(\Sigma)^{\mathrm{inv}}\,.
\end{equation}
In what follows we deal mainly with the algebra
$\Lambda(N)=\Lambda(\Sigma)^\mathrm{inv}$ of smooth polyvector
fields on $N$. In view of the regularity conditions (\ref{r1}) and
(\ref{r2}), the space $\Lambda(N)$ can also  be identified with
equivalence classes of projectible polyvector fields on $M$: Two
$k$-vectors $A, B\in \Lambda^k(M)$ are said to be
\textit{equivalent} ($A \sim B$) iff
\begin{equation}\label{equiv}
    A - B = T_a K^a + S^\alpha\wedge R_\alpha\,,
\end{equation}
for some $K^a\in \Lambda^k(M)$, $S^\alpha \in \Lambda^{k-1}(M)$,
and the $k$-vector  $A$ is said to be \textit{projectible} (onto
$N$)  if
\begin{equation}\label{prcon}
  [T_a, A]\sim 0 \,, \qquad   [R_\alpha, A]\sim 0\,.
\end{equation}
The polyvectors of the form $T_a K^a + S^\alpha\wedge R_\alpha$
will be called \textit{trivial}. These form a linear subspace
$\Lambda(M)^{\mathrm{triv}}=\{A\in \Lambda(M) | A\sim 0 \}$. Since
the trivial polyvectors are obviously projectible, we can write
\begin{equation}
 \Lambda(M)^{\mathrm{triv}}\subset \Lambda(M)^{\mathrm{pr}}\subset \Lambda(M)\,,
\end{equation}
and set
\begin{equation}\label{proj}
\Lambda(N)=\Lambda(M)^\mathrm{pr}/\Lambda(M)^{\mathrm{triv}}\,.
\end{equation}
It is the space
$\Lambda^0(N)=\Lambda^0(M)^\mathrm{pr}/\Lambda^0(M)^{\mathrm{triv}}$,
which is regarded as the \textit{space of physical observables} of
the generic gauge system. In the less formal physical terminology,
the physical observables are the gauge-invariant functions taken
modulo on-shell vanishing terms. Note that this definition does
not refer to any action functional or Hamiltonian structure.

Another relevant space is
$\Lambda^1(N)=\Lambda^1(M)^{\mathrm{pr}}/\Lambda^{1}(M)^{\mathrm{triv}}$.
Classically, the physical dynamics on $N$ is specified by a
projectible vector field $V\in \Lambda^1(M)^\mathrm{pr}$: The
evolution of a physical observable represented by $O\in
C^{\infty}(M)^{\mathrm{pr}}\equiv\Lambda^0(M)^\mathrm{pr}$ is
governed by the equation
\begin{equation}\label{vo}
    \dot O=[V,O]\,,
\end{equation}
where the overdot stands for the time derivative. If $O$ is a
projectible (trivial) function at the initial time moment, it will
remain a projectible (trivial) function in future as long as $V$
is a projectible vector field.

Bearing in mind the deformation quantization of the classical
gauge system (\ref{vo}), we have to equip the commutative algebra
of functions $C^{\infty}(N)\equiv\Lambda^0(N)$ with a Poisson
structure. For this end, we introduce a projectible bivector field
$P\in \Lambda^2(M)^\mathrm{pr}$ subject to the weak Jacobi
identity
\begin{equation}\label{pp}
    [P,P]\sim 0 \, ,
\end{equation}
and set
\begin{equation}\label{bracket}
   \{ A , B \} = P^{ij}\partial_i A\partial_j B \, ,
   \quad \forall A, B \in C^{\infty}(M) \,.
\end{equation}
Then for all projectible functions $O_1,O_2,O_3\in
C^{\infty}(M)^{\mathrm{pr}}$ and a trivial function $A\sim 0$ we
have
\begin{enumerate}
    \item  $\{O_1,A\}\sim 0$,
    \item $\{O_1,O_2\}\in C^{\infty}(M)^{\mathrm{pr}}$,
    \item $\{\{O_1,O_2\},O_3\}+c.p.(O_1,O_2,O_3)\sim 0$.
\end{enumerate}
Relations 1), 2) are just reformulations of the projectibility
conditions for $P$, whereas 3) is equivalent to the weak Jacobi
identity (\ref{pp}). Together relations 1),2),3) mean that the
brackets (\ref{bracket}) define the Poisson algebra structure  on
the space of physical observables. A projectible bivector $P$,
satisfying the above relations, will be called a \textit{weak
Poisson structure} associated to the constrained gauge algebra
(\ref{R}, \ref{rinv}).

A projectible vector field $V$ defines the Poisson vector field on
$N$ if and only if
\begin{equation}\label{vp}
    [V,P]\sim 0\,.
\end{equation}
In what follows we will refer to $V$ as a \textit{weak Poisson
vector field}. The condition (\ref{vp}) will be crucial upon
formulating a quantum counterpart of the classical equation of
motion (\ref{vo}).

Applying the Jacobi identity to the Schouten brackets (\ref{R}),
(\ref{rinv}), (\ref{pp}) and taking into account the regularity
conditions (\ref{r1}), (\ref{r2}), one can derive an infinite set
of higher structure relations underlying the constrained gauge
algebra endowed with a weak Poisson structure. In the next Section
we present a systematic BRST-like procedure for generating all
these relations. Also this BRST formalism will serve as basis for
the deformation quantization of the generic gauge system by means
of the formality theorem (see Sec. 4).

\section{BRST embedding of a generic gauge system}
The BRST embedding of any gauge system always starts with an
appropriate extension of the base manifold $M$ with ghost
variables. The conventional ways to introduce ghosts are different
for the Hamiltonian and Lagrangian BRST schemes. In this section
we suggest a uniform scheme for the BRST embedding of the generic
gauge system which does not involve Poisson structure or action.
Here we consider the constraints (\ref{T}) and gauge generators
(\ref{R}), (\ref{rinv}) as scalar functions and vector fields, so
the bundle of ghost variables is trivial. The generalization to
nontrivial vector bundles will be given in Sec. 5.

Consider a supermanifold $\mathcal{M}$ associated with the trivial
vector bundle $M \times\Pi\mathbb{R}^m\times \Pi\mathbb{R}^n
\rightarrow M$. Let $\eta_a$ and $c^\alpha$ denote linear
coordinates on the odd vector spaces $\Pi\mathbb{R}^m$ and
$\Pi\mathbb{R}^n$, respectively. If one regards the shell $\Sigma:
T_a=0$ as the surface of Hamiltonian first class constraints,
identifying the gauge generators $R_\alpha$ with the Hamiltonian
vector fields $R_a =\{T_a, \,\cdot\, \}$, then $c^\alpha$ and
$\eta_a$ will be the standard BFV ghosts and ghost momenta
respectively \footnote{For a Lagrangian gauge system, the
constraints $T_a=0$ have to be understood as equations of motion.
The vector fields $R$'s are then regarded as generators of the
gauge transformations, $c$'s are naturally identified with the
standard BV ghosts, and $\eta$'s play the role of BV antifields.
Note that in the presence of gauge symmetries the Lagrangian
equations of motion $T_a=0$ are necessarily reducible. In the
spirit of  usual ideas of the BRST theory, the ghost for ghost are
to be introduced in this case.  Here, we deal only with
irreducible case, although the generalization to reducible
constraints (which is necessary to describe accurately a
Lagrangian gauge theory) is a technical problem which will be
considered elsewhere.} (though in our approach they are no longer
assumed to be conjugated variables w.r.t. any Poisson bracket). To
construct the classical BRST embedding of the generic gauge it
would be sufficient, in principle, to work only with the
supermanifold $\mathcal{M}$, if the presence of a weak Poisson
structure was not crucial (see comments after Rel. (\ref{40})). To
trace the compatibility of the BRST embedding with the Poisson
algebra of physical observables we need, however, the further
extension of $\mathcal{M}$ to its odd cotangent bundle $\Pi
T^\ast\mathcal{M}$. There is an obvious isomorphism
\begin{equation}\label{dpr}
\Pi T^\ast\mathcal{M}\sim (\mathbb{R}^m)^\ast\times (
\mathbb{R}^n)^\ast\times \Pi \mathbb{R}^m\times \Pi
\mathbb{R}^n\times\Pi T^\ast M\,,
\end{equation}
where the asterisk means passing to a dual vector space. Let  $
\se^a$,  $\sc_\alpha$ and $\sx_i$ denote linear coordinates on
$(\mathbb{R}^m)^\ast$, $(\mathbb{R}^n)^\ast$ and $T^\ast_xM$,
respectively.

The supermanifold $\Pi T^\ast\mathcal{M}$ can be endowed with
various $\mathbb{Z}$-gradings in addition to the basic
$\mathbb{Z}_2$-grading (the Grassman parity). As usual, this is
achieved by prescribing certain integer degrees to the local
coordinates. It is convenient to arrange the information about
these gradings in a table:
\begin{equation}\label{table}
\begin{tabular}{|l|c|c|c|c|c|c|}
  \hline
           &\quad $x$\;\;\; & \quad$\sx$\;\;\; &\quad $c $\;\;\;&\quad $\sc$\;\;\; &\quad $\eta$\;\;\; &\quad $\se$\;\;\; \\
  \hline
  $\epsilon$=Grassman parity            & 0 & 1 & 1 & 0 & 1 & 0 \\
  \hline
  $\gh$ = ghost number                   & 0 & 1 & 1 & 0 & -1 & 2 \\
  \hline
  $\mathrm{Deg}$ = polyvector degree \;\;     & 0 & 1 & 0 & 1 & 0 & 1 \\
  \hline
  $\deg$ = resolution degree               & 0 & 0 & 0 & 1 & 1 & 0 \\
  \hline
\end{tabular}
\end{equation}
For brevity, the polyvector and resolution degrees will be
referred to as p- and r-degrees respectively. Splitting the
coordinates into \textit{fields} $\phi^A=(x^i,c^\alpha, \eta_a)$
and \textit{antifields} $\phi^\ast_A=(\sx_i, \sc_\alpha, \se^a)$,
we can write
\begin{equation}\label{}
    \gh(\phi^\ast_A)=-\gh(\phi_A)+1\,,\qquad
    \epsilon(\phi^\ast_A)=\epsilon(\phi^A)+1\,,\qquad \mathrm{Deg}
    (\phi^\ast_A)=1\,,\qquad \mathrm{Deg}(\phi^A)=0\,.
\end{equation}
Here the coordinates $\sc_\alpha$ are treated as formal variables
although they have the same parity and the ghost number as the
local coordinates $x^i$. The distinctions between $x$'s and
$\sc$'s are indicated by the auxiliary p- and r-degrees.

The odd cotangent bundle $\Pi T^\ast\mathcal{M}$ carries the
canonical antisymplectic structure.  The corresponding
anti-Poisson brackets of fields and antifields read
\begin{equation}\label{abr}
    (\phi^\ast_A,\phi^B)=\delta^B_A\,,\qquad (\phi^\ast_A,\phi^\ast_B)=0\,,\qquad
    (\phi^A, \phi^B)=0\,.
\end{equation}

Now we introduce a pair of master equations generating all the
structure relations underlying the constrained gauge algebra
(\ref{T}-\ref{rinv}) equipped with a weak Poisson structure $P$
and a weak Poisson vector field $V$. The equations read
\begin{equation}\label{me}
    (S,S)=0\,,\qquad (S,H)=0\,,
\end{equation}
where the functions $S$ and $H$ are subject to the following
grading and regularity conditions:
\begin{equation}\label{16}
\gh(S)=2\,,\qquad \epsilon(S)=0\,,\qquad  \mathrm{Deg}(S)> 0\,,
\end{equation}
\begin{equation}\label{17}
\gh(H)=1\,,\qquad \epsilon(H)=1\,,\qquad  \mathrm{Deg}(H)>0\,,
\end{equation}

\begin{equation}\label{18}
    \mathrm{rank}\left.\left(\frac{\partial^2
    S}{\partial\phi^A\partial\phi^\ast_B}\right)\right|_{dS=0}=(n, m)\,.
\end{equation}
Note that the last inequalities in (\ref{16}) and (\ref{17}) are
equivalent to vanishing of $S$ and $H$ on the Lagrangian surface
$\mathcal{M} \subset  \Pi T^* \mathcal{M} \, : \,
\,\phi^\ast_A=0$.

The general expansion for $S$, compatible with the grading
(\ref{16}), reads
\begin{equation}\label{S}
    \begin{array}{c}
     \displaystyle S =\; \se^a T_a + \sx_iR_\alpha^ic^\alpha +\sx_j\sx_iP^{ij}
      +\sc_\gamma(U^\gamma_{\alpha\beta}c^\beta c^\alpha + \sx_j\sx_iV^{\gamma ij}+ \sx_i c^\alpha W^{\gamma i}_\alpha+\se^a Y_a^\gamma)
      \\[3mm]
      +\eta_a ( \sx_iA^{ai}_{\alpha\beta}c^\beta c^\alpha+
      \sx_j\sx_iB^{aij}_\alpha c^\alpha+\sx_k\sx_j\sx_i D^{a
      ijk}+\se^bE^a_{b\alpha}c^\alpha+\sx_iF_{b}^{ai}\se^b)+\cdots\,,
    \end{array}
\end{equation}
where dots stand for the terms of r-degree $\geq 2$. The
coefficients  at the first three terms (all of r-degree $0$) are
identified respectively with the constraints (\ref{T}), the gauge
algebra generators (\ref{R}), (\ref{rinv}) and the  weak Poisson
bivector (\ref{pp}). Upon this identification the regularity
conditions (\ref{r1}), (\ref{r2}) are expressed by Rel.
(\ref{18}).

Up to the first order in r-degree a general expansion for $H$ is
given by
\begin{equation}\label{H}
    H=V^i\sx_i+ \sc_\alpha (W^\alpha_\beta c^\beta+\sx_i G^{\alpha i})+
    \eta_a(M^a_b\se^b+\sx_i\sx_j L^{aji}+\sx_i N_\alpha^{ai}c^\alpha)+\cdots\,,
\end{equation}
where the first term is identified with the projectible vector
field entering the equation of motion (\ref{vo}).

The existence of solutions to the master equations (\ref{me}) is
proved by the standard tools of homological perturbation theory
\cite{ht}, \cite{St}.  Below, we sketch the proof for $S$; for
$H$, the existence theorem is proved in a similar way.

Substituting the general expansion
\begin{equation}\label{}
    S=\sum_{n=0}^\infty S_n\,,\qquad \deg(S_n)=n\,,
\end{equation}
in the first equation of (\ref{me}), we arrive at a chain of
equations of the form
\begin{equation}\label{sn}
    \delta S_{n+1}=K_n(S_0,...,S_n) \,, \qquad \deg(K_n)=n\,,
\end{equation}
where
\begin{equation}\label{delta}
    \delta =T_a\frac{\partial}{\partial \eta_a}+\sx_iR^i_\alpha\frac{\partial}{\partial\sc_\alpha}
\end{equation}
is a nilpotent operator decreasing the r-degree by one unit,
\begin{equation}\label{}
\delta^2=0\,,\qquad\deg(\delta)=-1\,,
\end{equation}
and $K_n$ involves the antibrackets of the $S$'s of lower order.
Let $\mathcal{H}(\delta)=\bigoplus \mathcal{H}_n(\delta)$ denote
the corresponding cohomology group graded naturally by r-degree.
>From the regularity conditions (\ref{r1}), (\ref{r2}) it follows
that
\begin{equation}\label{Hom}
\mathcal{H}_n(\delta)=0\, \quad \mathrm{for} \quad n>0\,.
\end{equation}
Expanding the first equation $\delta S_1=K_0(S_0)$ in the ghost
variables one recovers the involution relations (\ref{R}),
(\ref{rinv}), the projectibility conditions (\ref{prcon}) for the
weak Poisson bivector $P$,
\begin{equation}\label{}
    [T_a,P]=-Y_a^\alpha \wedge R_\alpha -T_b F^b_a\,,\quad [R_\alpha, P]=W_\alpha^\beta \wedge R_\beta-T_aB^a_\alpha\,,
\end{equation}
and the weak Jacobi identity (\ref{pp}),
\begin{equation}\label{}
\begin{array}{c}
    \frac12[P,P]=-V^\alpha\wedge R_\alpha -T^aD_a\,.
    \end{array}
\end{equation}
Now the existence of $n$-order solution can be proved by
induction. Indeed, from the Jacobi identity $(S,(S,S))\equiv 0$ it
follows that the r.h.s. of $n$-th equation is $\delta$-closed,
provided that the previous $(n-1)$-th equations are satisfied. In
view of (\ref{Hom}), this $\delta$-closed expression is exact,
i.e. some $S_{n+1}$ exists obeying (\ref{sn}). Using the induction
on $n$ one can also see that $S_{n+1}|_{\phi^\ast=0}=0$, as the
operator $\delta$ does not change p-degree, while the antibrackets
decrease the degree by one.

Thus, the master equation $(S,S)=0$ is proved to be soluble for an
arbitrary constrained gauge algebra and a weak Poisson structure.

Mention that neither $S$ nor $H$ are uniquely determined by the
master equations (\ref{me}).
 The reason is that one can always make a canonical transformation $f: \Pi T^\ast\mathcal{M}
\rightarrow \Pi T^\ast\mathcal{M}$ (i.e. a diffeomorphism
respecting the canonical antibrackets (\ref{abr})) and get new
solutions $S'=f^\ast (S)$ and $H'=f^\ast (H)$ to the same
equations. Consider, for instance, a finite anti-canonical
transformation
\begin{equation}\label{ctr}
    F' = e^{(G,\,\cdot\,)}F=\sum_{n=0}^\infty \frac1{n!}\underbrace{(G,(G,\cdots(G}_{n} ,F)\cdots)\,,\qquad \forall
    \,F\in C^\infty(\Pi T^\ast\mathcal{M})\,.
\end{equation}
generated by the odd, ghost number 1 function
\begin{equation}\label{}
    G= c^\alpha A_\alpha^\beta(x)\sc_\beta+\eta_a B^a_b(x)\se^b + \eta_a c^\alpha Z^{ai}_\alpha(x)\sx_i\,.
\end{equation}
By applying this transformation to $S$ and $H$, one redefines the
structure functions entering the expansions (\ref{S}) and
(\ref{H}), including those ones which are identified with the
original constraints and gauge generators:
\begin{equation}\label{tr}
T_a\,\rightarrow T'_a=\tilde{A}_a^bT_b\,,\qquad    R_\alpha
\,\rightarrow\, R'_\alpha
=\tilde{B}^\beta_\alpha(R_\beta+T_a\tilde{Z}^a_{\beta} )\,,
\end{equation}
where
\begin{equation}\label{}
    \tilde{A}=e^A\,,\qquad \tilde{B}=e^B\,,\qquad \tilde{Z}^i=\int_0^1 e^{tA}Z^ie^{-tB}dt\,.
\end{equation}
Relations (\ref{tr}) reflect an inherent ambiguity in the
definition of the constraints and gauge algebra generators. The
invariant geometrical meaning can be assigned only to
 the shell $\Sigma$ (a particular choice of the constraints $T_a$
is not significant if they have the same zero locus), and the
gauge foliation $\mathcal{F}(\Sigma)$ (various choices of the
vector fields $R_\alpha$ are all equivalent if they define the
same on-shell integrable distribution). By adding to $G$ the term
$W^i(x)\sx_i$, $W^i\partial_i$ being a complete vector field on
$M$, one generates just a diffeomorphism of the original manifold
$M$. The other possible terms that one may add to $G$ do not
contribute to the transformations (\ref{tr}). A deeper geometric
insight into the nature of these transformations will be given in
Sec. \ref{gen}, where the  $T$'s and $R$'s are considered to be
sections of appropriate vector bundles with connection rather than
sets of functions  and vector fields.

Consider another interpretation for the master equations
(\ref{me}) in terms of polyvector algebra. It is based on an
obvious identification of the odd Poisson algebra on $\Pi
T^\ast\mathcal{M}$ with the exterior algebra of polyvector fields
$\Lambda(\mathcal{M})=\bigoplus\Lambda^n(\mathcal{M})$ on
$\mathcal{M}$ endowed with the Schouten brackets. Upon this
interpretation the anti-fields $\phi^\ast_A$ play the role of the
natural frame $\partial_A$ in the fibers of the odd tangent bundle
$\Pi T\mathcal{M}$ and the rank of a polyvector coincides with the
p-degree of the corresponding superfunction. Each homogeneous
subspace $\Lambda^n(\mathcal{M})=\bigoplus
\Lambda^n_m(\mathcal{M})$ is further graded by the ghost number
$m=0,1,...$ .

The expansions of $S$ and $H$ w.r.t. the p-degree read
\begin{equation}\label{sqp}
\begin{array}{c}
\displaystyle    S=\sum_{n=1}^\infty S^{n} =
    Q^A\phi^\ast_A+\Pi^{AB}\phi^\ast_B\phi^\ast_A+\Psi^{ABC}\phi^\ast_C\phi^\ast_B\phi^\ast_A+\cdots \,,\\[3mm]
\displaystyle    H=\sum_{n=1}^\infty H^n=
\Gamma^A\phi^\ast_A+\Xi^{AB}\phi_A^\ast\phi_B^\ast+\cdots\, ,
    \end{array}
\end{equation}
where $\mathrm{Deg}(S^n)=\mathrm{Deg}(H^n)=n$.  Note that the
contributions of $0$-vectors to $S$ and $H$ are prohibited by
(\ref{16}), (\ref{17}).

To simplify notation, we use the same letter $F$ for a
superfunction $F(\phi^A,\phi^\ast_B) \in C^{\infty}(\Pi
T^\ast\mathcal{M})$ and for the corresponding polyvector field
$F(\phi^A,\partial_B)\in \Lambda(\mathcal{M})$. Then the master
equations on $S$ and $H$ yield two sequences of relations for the
homogeneous polyvectors:
\begin{equation}\label{r}
[S,S]=0\;\Rightarrow\;  [Q,Q]=0\,,\qquad [Q, \Pi]=0\,,\qquad
[\Pi,\Pi]=
    -2[Q, \Psi]\,,\quad\cdots\,,
\end{equation}
\begin{equation}\label{rr}
    [S,H]=0\;\Rightarrow\; [Q,\Gamma]=0\,,\qquad [\Gamma,\Pi]=[Q,\Xi]\,,\quad\cdots\,.
\end{equation}
The first relation in (\ref{r}) means that
\begin{equation}\label{40}
    Q=T_a\frac{\partial}{\partial \eta_a}+c^\alpha R_\alpha^i\frac{\partial}{\partial
    x^i}+c^\beta c^\alpha U^\gamma_{\alpha\beta}\frac{\partial}{\partial
    c^\gamma}+\eta_aA^{ai}_{\alpha\beta}c^\beta c^\alpha\frac{\partial}{\partial
    x^i}+\cdots
\end{equation}
is an integrable odd vector field on $\mathcal{M}$ of  ghost
number 1. In mathematics, such a value is known as a homological
vector field (see e.g. \cite{vain}). From the physical viewpoint,
it is $Q$ which serves as a BRST operator, generating classical
BRST transformations on $\mathcal{M}$ (i.e. on the original
manifold $M$ extended with the ghosts $c$'s and $\eta$'s). To
construct $Q$, one could directly solve the nilpotency condition
$[Q,Q]=0$ without resorting to antifields $\phi^\ast_A$.
Obviously, $Q$ contains all the information about the constrained
gauge algebra (\ref{T}, \ref{R}, \ref{rinv}) defined independently
of the weak Poisson structure $P$. The classical evolution
(\ref{vo}) is generically described by the weak Poisson vector
field $V$. If one identifies the base manifold $M$ with the space
of trajectories of a gauge system, the constraints $T_a=0$ can be
interpreted as equations of motion; in so doing, the vector field
$V$ can be regarded as the generator of a global symmetry of the
system. The configuration space dynamics can be described, in this
sense, only by the constraints, i.e. without independent equations
of motion. In this context, the BRST variation $Q\eta_a$ in the
sector of variables of ghost number -1 can be identified with the
equations of motion, which may be not necessarily Lagrangian, as
it was subtly noticed in Ref. \cite{BGST}.

The adjoint action of $Q$ gives rise to a nilpotent
differentiation $D:\Lambda^n_m(\mathcal{M})\rightarrow
\Lambda_{m+1}^n(\mathcal{M})$:
\begin{equation}\label{}
\begin{array}{rl}
    DA=[Q,A]\,,\qquad& \forall\, A\in\Lambda(\mathcal{M})\,,\\[3mm]D^2=0\,,
    \qquad &
    \gh(D)=1\,.
    \end{array}
\end{equation}
The corresponding cohomology group can be decomposed
$\mathcal{H}(D)=\bigoplus \mathcal{H}^n_m(D)$ according to the
bi-grading by the p-degree and the ghost number.

The weak Poisson structure enters the expansion (\ref{sqp})
through the bivector $\Pi$. The latter is a $D$-cocycle as well as
the vector $\Gamma$. The third relation in (\ref{r}) means that
$\Pi$ is a Poisson bivector up to a $D$-coboundary. Similarly, the
second relation in (\ref{rr}) characterizes $\Gamma$ as weak
Poisson vector field on $\mathcal{M}$.

The higher polyvectors entering $S$ and $H$ contribute to various
compatibility conditions between $Q$, $\Pi$ and $\Gamma$, which
result from the Jacobi identity for the Schouten brackets and the
triviality of the $D$-cohomology groups $\mathcal{H}^n_m(D)$ with
$m<n$. The last fact can be seen as follows: According to
(\ref{table}), $\mathrm{Deg} (A)>\gh (A)$ implies $\deg(A)>0$, but
$D$ is the perturbation of $\delta$ by terms of higher r-degree,
and the triviality of the $D$-cocycle $A$ follows immediately from
acyclicity of $\delta$ in strictly positive r-degree (\ref{Hom}).

Let us show that the space of physical observables
$C^{\infty}(N)=C^\infty(M)^{\mathrm{pr}}/C^\infty(M)^{\mathrm{triv}}$
is isomorphic to $\mathcal{H}^0_0(D)$. Substituting a general
expansion
\begin{equation}\label{exp1}
    \Lambda^0_0(\mathcal{M})\ni F=\sum_{n=0}^\infty F_n(\phi) =f(x)+ \eta_aV^a_\alpha(x) c^\alpha+\cdots\,,\qquad
    \deg(F_n)=n\,,
\end{equation}
into the $D$-closedness condition
\begin{equation}\label{Dcl}
    DF=[Q,F]=(S,F)|_{\phi^\ast=0}=0\,,
\end{equation}
we get a sequence of equations of the form
\begin{equation}\label{AB}
 \delta F_{n+1}=B_n(F_0,...,F_n)\,,\qquad \deg(B_n)=n\,,
\end{equation}
where  the nilpotent differential (\ref{delta}) is now truncated
to $\delta=T_a\partial/\partial
 \eta_a$, which is the usual Koszul-Tate differential associated to the constraint surface $\Sigma$.
One may check that the first of the equations (\ref{AB}) is
nothing but the projectibility condition for the function $f(x)$:
\begin{equation}\label{}
    [R_\alpha,f]=V_\alpha^aT_a\,.
\end{equation}
Using the identity $D^2F\equiv 0$, one can see that the r.h.s. of
the $n$-th equation is $\delta$-closed provided all the previous
equations are satisfied. Since $\mathcal{H}_n(\delta)=0$ for all
$n>0$, we conclude that (i) any projectible function $f\in
C^\infty(M)^{\mathrm{pr}}$ is lifted to a $D$-cocycle  $F\in
C^\infty(\mathcal{M})$ and (ii) any two equivalent  (in the sense
of (\ref{equiv})) functions $f_1,f_2\in C^\infty(M)^{\mathrm{pr}}$
belong to the same cohomology class upon the lift, i.e.
\begin{equation}\label{}
    F_1-F_2=DK\,,\qquad K=K^a\eta_a+\cdots\,.
\end{equation}
This establishes the isomorphism $C^{\infty}(N)\simeq
\mathcal{H}^0_0(D)$.

Now consider the following brackets on $C^\infty(\mathcal{M})$:
\begin{equation}\label{br}
    \{A,B\}=(A,(S,B))|_{\phi^\ast=0}=[A,[\Pi,B]]\,.
\end{equation}From the third  relation in (\ref{r}) it follows that for any
$A,B,C\in C^{\infty}(\mathcal{M})$ we have
\begin{equation}\label{}
\begin{array}{c}
   (-1)^{\epsilon(A)\epsilon(C)} \{\{A,B\},C\}+\cycle (A,B,C)= D\Psi(dA,dB,dC)\\[3mm]
    +\Psi(dDA,dB,dC)+(-1)^{\epsilon(A)}\Psi(dA,dDB,dC)+(-1)^{\epsilon(A)+\epsilon(B)}\Psi(dA,dB,dDC)\,.
\end{array}
\end{equation}
In other words, the brackets (\ref{br}) induce a Poisson structure
on the cohomology group $\mathcal{H}^0(D)$ considered as a
supercommutative algebra. In particular, the physical observables
form a closed Poisson subalgebra $\mathcal{H}^0_0(D)$.

Finally, the evolution of a physical observable $[A]\in
\mathcal{H}_0^0(D)$ represented by a $D$-cocycle $A$ is determined
by the equation
\begin{equation}\label{da}
    \dot A=[\Gamma, A]=(H,A)|_{\phi^\ast=0}\,.
\end{equation}
In view of Rels. (\ref{rr}) it defines a one-parameter
automorphism of the Poisson algebra $\mathcal{H}_0^0(D)$.

We may now summarize that the spaces of physical observables
$C^{\infty}(N)$, physical evolutions $\Lambda^1(N)$ and the
Poisson structures on $N$ are in one-to-one correspondence
respectively with the cohomology groups $\mathcal{H}_0^0(D)$,
$\mathcal{H}_0^1(D)$ and $\mathcal{H}_0^2(D)$ of the BRST operator
$D$.

\section{Quantization by means of formality theorem}

The aim of this section is to perform the deformation quantization
of the generic gauge system endowed with the weak Poisson brackets
(\ref{br}). For this purpose we apply the covariant version
\cite{CFT}, \cite{Dolg} of Kontsevich's formality theorem
\cite{Kontsevich}, which states the existence of quasi-isomorphism
\begin{equation}\label{}
F: \Lambda(\mathcal{M})\rightsquigarrow D(\mathcal{M})
\end{equation}
between two differential graded Lie algebras \footnote{More
precisely, here we mean the generalization of the formality
theorem to the case of supermanifolds. No such a generalization
has been published yet, although its existence was never doubted,
as far as we could know.}: the algebra $(\Lambda(\mathcal{M}), [
\, \cdot \,, \, \cdot \, ],0)$ of polyvector fields equipped with
the Schouten bracket and trivial differential, and the algebra
$(D(\mathcal{M}), [\,\cdot\,,\cdot\,]_{\cal {G}}, \delta)$ of
polydifferential operators $D(\mathcal{M})=\bigoplus
D^k(\mathcal{M})$ on $\mathcal{M}$, where $D^k(\mathcal{M})$ is
the space of differential operators acting on $k$ functions,
 $[\,\cdot\,,\,\cdot\,]_{\cal {G}}$ is the Gerstenhaber bracket and
$\delta: D^k(\mathcal{M})\rightarrow D^{k+1}(\mathcal{M})$ is the
Hochschild differential. Mention that if $\Lambda(\mathcal{M})$
and $D(\mathcal{M})$ are regarded as differential graded Lie
algebras, the multiplicative gradings are shifted by -1.

To any collection of polyvector fields $A_1,...,A_n$ of degrees
$k_1,...,k_n$ the map $F$ assigns a polydifferential operator
$F_n(A_1,...,A_n)\in D^m(\mathcal{M})$ acting on
$m=2-2n+\sum_{i=1}^n k_i$ functions such that the following
semi-infinite sequence of quadratic relations is satisfied ($n\geq
1$):
\begin{equation}\label{F}
\begin{array}{r}
\displaystyle\delta F_n(A_1,...,A_n)=\frac12 \sum_{k,l\geq 1,\,
k+l=n}\frac{1}{k!l!}\sum_{\sigma\in S_n} \pm [
F_k(A_{\sigma(1)},...,A_{\sigma(k)}),
F_l(A_{\sigma(k+1)},...,A_{\sigma(k+l)})]_{\cal {G}}\\[5mm]
\displaystyle +\sum_{i\neq j} \pm F_{n-1}([A_i,A_j],A_1...,
\hat{A}_i,...,\hat{A}_j,...,A_n)\,.
\end{array}
\end{equation}
Here $S_n$ is the group of permutations of $n$ letters and the
caret denotes omission. The rule for determining signs in the
above formula is rather intricate. For example, the sign factor in
the first sum depends on the permutation $\sigma$, on degrees of
$A_i$ and on the numbers $k$ and $l$.

Mention also the symmetry property
\begin{equation}\label{sym}
    F_n(A_1,...,A_i,...,A_j,...,A_n)=(-1)^{\epsilon(A_i)\epsilon(A_j)}F_n(A_1,...,A_j,...,A_i,...,A_n)\,,
\end{equation}
and the ``boundary'' condition
\begin{equation}\label{bc}
    F_1(A)(f_1,...,f_m)=(-1)^m A(df_1,...,df_m)\,,\qquad \forall A\in \Lambda^m(\mathcal{M})\,.
\end{equation}

Let us now apply the identities (\ref{F}) to $F_{n}(S,..., S)$. In
view of the master equation
\begin{equation}\label{me1}
[S,S]=0\,,
\end{equation}
the sum in the second line of (\ref{F}) vanishes and the whole
sequence of relations can be combined in a single equation
\begin{equation}\label{gme}
 [\hat{S},\hat{S}]_{\cal {G}}=0
\end{equation}
for the inhomogeneous polydifferential operator
\begin{equation}\label{}
D(\mathcal{M})[[\hbar]]\ni\hat{S}=\sum_{n=0}^\infty\frac{\hbar^n}{n!}F_n(S,...,S)\,,
\end{equation}
$\hbar$ being formal parameter. Here we use the fact that the
Hochcshild differential $\delta$ is an inner derivation of the
Gerstenhaber algebra, namely,
\begin{equation}\label{}
    \delta = [ F_0,\,\cdot\,]_{\cal {G}}\,,
\end{equation}
where $F_0\in D^2(\mathcal{M})$ is the multiplication operator,
$F_0(f,g)=fg$, $\forall f,g\in C^{\infty}(\mathcal{M})$. Expanding
$\hat{S}$ in the sum of homogeneous polydifferential operators
\begin{equation}\label{hs}
\begin{array}{c}
   \displaystyle \hat{S}=\sum_{n=0}^\infty \hat{S}^n =\hat{A}+\hat{Q}+\hat{\Pi}+\hat{\Psi}+\hat{S}^4+\cdots\,,\\[5mm]
\displaystyle     D^n(\mathcal{M})[[\hbar]]\ni \hat{S}^n =
\sum_{k=0}^\infty \frac{\hbar^k}{k!}\sum_{l_1+\cdots + l_k
=n+2(k-1)}
    F_k(S^{l_1},...,S^{l_k})\,.
\end{array}
\end{equation}
and taking into account (\ref{sym}) and (\ref{bc}), we can write
\begin{equation}\label{qcl}
\begin{array}{ll}
\displaystyle \hat{A}= \frac{\hbar^2}2 F_2(Q,Q)+\frac{\hbar^3}2
F_3(Q,Q, \Pi) + O(\hbar^4) \,,\quad&
    \hat{Q}(f) = -\hbar Q( df) + O(\hbar^2)\,,\\[3mm]
   \displaystyle  \hat{\Pi} (f, g)= fg +\frac\hbar 2\Pi(df,dg)+ O(\hbar^2)\,,&
    \hat{\Psi}(f,g,h) = -\hbar \Psi(df,dg,dh)+O(\hbar^2)\,,
\end{array}
\end{equation}
for $f,g,h \in C^{\infty}(\mathcal{M})$. Now substituting
(\ref{hs}) into the ``Gerstenhaber master equation'' (\ref{gme}),
we get a sequence of relations, the first four
 of which are
\begin{equation}\label{rel}
 \begin{array}{rlrl}
    i) & [ \hat{A},\hat{Q}]_{\cal {G}}=0\,, & iii)&[ \hat{Q},\hat{\Pi}]_{\cal {G}} =-[ \hat{A},\hat{\Psi}]_{\cal {G}}\,,\\[3mm]
    ii)&[ \hat{Q},\hat{Q}]_{\cal {G}}=-2[ \hat{A}, \hat{\Pi}]_{\cal {G}}\,,\quad& iv)&[ \hat{\Pi},\hat{\Pi}]_{\cal {G}} = -2[ \hat{Q},\hat{\Psi}]_{\cal {G}}
    -2[ \hat{A},\hat{S}^4]_{\cal {G}}\,.
    \end{array}
\end{equation}
Let us suppose $\hat{A}=0$, for a while. Then Rels. (\ref{rel})
can be interpreted as follows:

$(ii)$ The homological vector field $Q$ is lifted to the formal
differential operator
$$\hat{Q}: C^{\infty}(\mathcal{M})[[\hbar]]\rightarrow C^{\infty}(\mathcal{M})[[\hbar]]\,.$$
The latter is odd and nilpotent,
\begin{equation}
 \hat{Q}^2=0\,.
 \end{equation}
Let us denote by $\mathcal{H}(\hat{Q})=\bigoplus
\mathcal{H}^n(\hat{Q})=\ker \hat{Q}/{\mathrm{im}\hat{Q}}$ the
corresponding cohomology group, graded naturally by the ghost
number.

$(iv)$ The bidifferential operator
$$\hat{\Pi}:C^\infty(\mathcal{M})[[\hbar]]\otimes
 C^{\infty}(\mathcal{M})[[\hbar]]\rightarrow
C^{\infty}(\mathcal{M})[[\hbar]]$$ defines the $\ast$-product
\begin{equation}\label{}
    f\ast g =\hat{\Pi}(f,g)= fg+\frac\hbar 2\{f,g\} + O(\hbar^2)\,,
\end{equation}
which is a formal deformation of the pointwise product of
functions in the ``direction'' of the weak Poisson bracket
(\ref{br}). This $\ast$-product possesses the property of weak
associativity:
\begin{equation}\label{wa}
\begin{array}{c}
    (f\ast g)\ast h -f\ast(g\ast h)=\hat{Q}(\hat{\Psi}(f,g,h)) \\[3mm]
    +\hat{\Psi}(\hat{Q}(f),g,h)+(-1)^{\epsilon(f)}\hat{\Psi}(f,\hat{Q}(g),h)+
    (-1)^{\epsilon(f)+\epsilon(g)}\hat{\Psi}(f,g,\hat{Q}(h))\,,
    \end{array}
\end{equation}
for all $f,g,h\in C^{\infty}(\mathcal{M})[[\hbar]]$.

$(iii)$ $\hat{Q}$ differentiates the $\ast$-product:
\begin{equation}\label{dif}
{\hat{Q}}(f\ast g)=\hat{Q}(f)\ast g+(-1)^{\epsilon(f)}f\ast
\hat{Q}(g)\,.
\end{equation}

Together, Rels. (\ref{wa}), (\ref{dif}) suggest that the weakly
associative $\ast$-product on $C^\infty(\mathcal{M})[[\hbar]]$
induces an \textit{associative} $\ast$-product on the cohomology
group $\mathcal{H}(\hat{Q})$. In particular,
$\mathcal{H}^0(\hat{Q})$ - the group of the $\hat{Q}$-cohomology
of ghost number zero - inherits the structure of
$\ast$-subalgebra in $H(\hat{Q})$. It is this subalgebra which is
identified with the algebra of \textit{quantum observables}.

When $\hat{A}\neq 0$, the above interpretation of Rels. $(ii),
(iii), (iv)$ becomes incorrect: The nonzero value of  $\hat{A}$
may break the nilpotency of the operator $\hat{Q}$ as well as the
property of the weak associativity (\ref{wa}) and the Liebniz rule
(\ref{dif}). One may thus regard the function  $\hat{A}$, having
no classical counterpart, as a \textit{quantum anomaly}. By
definition $ \gh(\hat{A})=2$, and hence it is expanded as
\begin{equation}\label{}
\hat{A}=c^\alpha c^\beta f_{\alpha\beta}(x,\hbar)+c^\alpha c^\beta
c^\gamma f_{\alpha\beta\gamma}^a
    (x,\hbar)\eta_a+\cdots\,.
\end{equation}
Actually, the function $F_2(Q,Q)$ vanishes identically
\cite{Kontsevich}, so the quantum anomaly starts with $\hbar^3$.
The last fact agrees well with the absence of one-loop anomalies
for the Weyl symbols of operators.

Let us now expand $\hat{A}$ in powers of $\hbar$:
\begin{equation}\label{An}
    \hat{A}=\hbar^n {A}^{(n)}+\hbar^{n+1} {A}^{(n+1)}+\cdots\,,\qquad n\geq 3\,.
\end{equation}
If the leading term $A^{(n)}$ is not equal to zero identically, we
refer to $n$ as the order of the anomaly $\hat{A}$. From the first
relation  in (\ref{rel}) follows that ${A}^{(n)}$ is a
$D$-cocycle:
\begin{equation}\label{DA}
    DA^{(n)}\equiv[Q,A^{(n)}]=0\;\Rightarrow\;[A^{(n)}]\in \mathcal{H}^0_2(D)\,.
\end{equation}
The anomaly $\hat{A}$ is said to be trivial at the lowest order
$n$, if $[A^{(n)}]=0$. In this case, there exists a function $B$
such that $A^{(n)}=DB$, and one can remove the leading  term
$A^{(n)}$ of the anomaly by an appropriate canonical
transformation. Namely, applying (\ref{ctr}) to $S$ with
$G=\hbar^{n-1}B$, we get a new solution
$S'=S-\hbar^{n-1}A^{(n)}+\cdots$ to the classical master equation
(\ref{me1}) and the corresponding solution $\hat{S}'$ to the
Gerstenhaber master equation (\ref{gme}); but the order of the
anomaly term $\hat{A}'$ is now equal to or grater than $n+1$.

If $A'^{(n+1)}$ happens to be a $D$-coboundary again, we may
repeat the above procedure once again and get an anomaly of order
$\geq n+2$. Doing this inductively on $n$, we can shift the
anomaly to higher orders in $\hbar$ until we face with a
nontrivial $D$-cocycle. In this case the procedure stops and we
get a genuine quantum anomaly, which cannot be removed by the
canonical transform (\ref{ctr}). There is, however, a hypothetical
possibility to cancel this anomaly out by means of a nontrivial
deformation of $S$ (i.e. a deformation which cannot be induced by
a canonical transform):
\begin{equation}
S\rightarrow S(\hbar)=S+\hbar
S_1+\cdots\,,\qquad[S(\hbar),S(\hbar)]=0\,.
\end{equation}
The existence of such a deformation and a concrete recipe for its
construction require a separate study going beyond the scope of
the present paper.

Now let us formulate a quantum counterpart for the classical
equation of motion (\ref{da}). In order to do this consider the
inhomogeneous polyvector $W=H+S$. The classical master equations
(\ref{me}) are obviously equivalent to the following one:
\begin{equation}\label{}
    [W,W]=0\,.
\end{equation}
Proceeding analogously to the previous case with $S$, we apply the
formality map  to $W$ and get inhomogeneous polydifferential
operator $\hat{W}\in D(\mathcal{M})[[\hbar]]$ satisfying the
Gerstenhaber master equation
\begin{equation}\label{ww}
    [\hat{W},\hat{W}]_{\mathcal{G}}=0\,.
\end{equation}
Since $\epsilon(H)=1$, $\hat{W}$ is at most linear in $H$, as is
clear from (\ref{sym}). Thus $\hat{W}=\hat{H}+\hat{S}$, where
\begin{equation}\label{}
\hat{H}=\sum_{n=0}^\infty\frac{\hbar^{n+1}}{n!}F_{n+1}(H,{S,...,S})
\end{equation}
Expanding the formal polydifferential operator $\hat{H}$ in its
homogeneous components, we get
\begin{equation}\label{hexpan}
    \hat{H}=\sum_{n=0}^\infty H^n=\hat{A}'+\hat{\Gamma}+\hat{\Xi}+\cdots\,,
\end{equation}
where
\begin{equation}\label{aexp}
\begin{array}{l}
    \hat{A}'=\hbar^2 F_2(Q,\Gamma) +\hbar^3F_3(Q,\Gamma,\Pi)+O(\hbar^4)\,,\\[3mm]
    \hat{H}(f)=-\hbar \Gamma(df) +O(\hbar^2)\,,\\[3mm]
   \hat{\Xi}(f,g)=\hbar\Xi(df,dg)+O(\hbar^2)\,,
\end{array}
\end{equation}
and $f,g\in C^\infty(\mathcal{M})$. Let us suppose that the
anomaly $\hat{A}$ equals to zero. Then substituting (\ref{hexpan})
into (\ref{ww}), we get
\begin{equation}\label{hrel}
\begin{array}{ll}
i) & [\hat{Q},\hat{A}']_{\mathcal{G}}=0\,,\\[3mm]
ii)& [\hat{Q},\hat{\Gamma}]_{\mathcal{G}}=-[\hat{A}',\hat{\Pi}]_{\mathcal{G}}\,,\\[3mm]
iii)&[\hat{\Gamma},\hat{\Pi}]_{\mathcal{G}}=[\hat{Q},\hat{\Xi}]_{\mathcal{G}}+
[\hat{\Psi},\hat{A}']_{\mathcal{G}}\,.
    \end{array}
\end{equation}
By definition, $\hat{A}'$ is a scalar function of ghost number 1.
Consequently,
\begin{equation}\label{}
    \hat{A}'=c^\alpha f_\alpha(x,\hbar)+c^\alpha c^\beta f_{\alpha\beta}^a\eta_a+\cdots\,.
\end{equation}
Having no classical counterpart, it represents one more quantum
anomaly related to the dynamics of the constrained gauge system.
Again, $F_2(Q,\Gamma)\equiv 0$, and the expansion (\ref{aexp}) for
$\hat{A}'$ starts actually with $\hbar^3$:
\begin{equation}\label{}
    \hat{A}'=\hbar^n A'{}^{(n)}+\hbar^{n+1}A'{}^{(n+1)}+\cdots\,,\qquad n\geq 3\,.
\end{equation}
>From the first relation of (\ref{hrel}) it then follows that the
leading term $A'{}^{(n)}$ is a $D$-cocycle, i.e. $[A'{}^{(n)}]\in
\mathcal{H}^0_1(D)$. Again, if $A'^{(n)}=DB'$ one can remove this
term by an appropriate canonical transform in perfect analogy to
the case of $\hat{A}$.

If now $\hat{A}'=0$, then the last two relations of (\ref{hrel})
suggest that the operator
$$\hat{\Gamma}: C^{\infty}(\mathcal{M})[[\hbar]]\rightarrow C^{\infty}(\mathcal{M})[[\hbar]]$$
commutes with $\hat{Q}$ and differentiates the $\ast$-product up
to homotopy,
\begin{equation}\label{}
    \hat{\Gamma} (f\ast g)-(\hat{\Gamma} f)\ast g-f\ast(\hat{\Gamma}g)= \hat{Q}(\hat{\Xi}(f,g))+\hat{\Xi}(\hat{Q}(f),g)
    +(-1)^{\epsilon(f)}\hat{\Xi}(f,\hat{Q}(g))\,.
\end{equation}
As a consequence, $\hat{\Gamma}$ induces a differentiation of the
algebra of physical  observables $\mathcal{H}^0_0(\hat{Q})$.

Now if both the anomalies $\hat{A}$ and $\hat{A}'$ vanish, the
quantum evolution is governed by the equation
\begin{equation}\label{}
    \dot O = \hat{\Gamma}O\,,
\end{equation}
where $O$ is a $D$-cocycle representing an observable $[O]\in
\mathcal{H}^0_0(\hat{Q})$.

In  field theory, one should be cautious about  the aforementioned
possibilities of removing quantum anomalies because of possible
divergences appearing when the polydifferential operators are
applied to local functionals.  Even when
$\mathcal{H}^0_1(D)=\mathcal{H}^0_2(D)=0$ the formally BRST-exact
contributions can diverge resulting in anomalies, whereas after a
regularization they may happen to be no longer BRST-closed.

\section{Weak Lie algebroids  and a topological sigma-model}
\label{gen} The concept of constrained gauge algebra, exposed in
Sec.2, admits a straightforward generalization to the case when
the constraints $T_a$ and the gauge algebra generators $R_\alpha$
are considered to be sections of some vector bundles over $M$
rather than sets of functions and vector fields. Namely, we may
set
$$T=T_a e^a \in \Gamma(E_1)\,,\qquad R=R_\alpha^i
e^\alpha\otimes\partial_i\in \Gamma(E_2\otimes TM)\,,$$ where
$\{e^a\}$ and $\{e^\alpha\}$ are local frames of two vector
bundles $E_1\rightarrow M$ and $E_2\rightarrow M$, and
${\partial_i=\partial/\partial x^i}$ is the natural frame in the
tangent bundle of the base $M$.

Notice that all the structure relations (\ref{T}-\ref{rinv}) of
the constrained gauge algebra as well as  the projectibility
conditions (\ref{prcon}) are form-invariant under the bundle
automorphisms.  The shell $\Sigma$ is now identified with the zero
locus of the section $T$; the latter is supposed to intersect the
base $M$ transversally in order for $\Sigma\subset M$ to be a
smooth submanifold (cf. (\ref{r1})). The section $R$ defines (and
is defined by) a bundle homomorphism $R: E^\ast_2\rightarrow TM$.
The regularity condition (\ref{r2}) is then equivalent to the
injectivity of $R|_\Sigma$. Rel. (\ref{R}) suggests that
$\mathrm{im}(R|_\Sigma) \subset T \Sigma \subset TM|_\Sigma$,
whereas Rel. (\ref{rinv}) identifies $\mathrm{im} (R|_\Sigma) $ as
an integrable distribution on $\Sigma$. In other words,
$E_2|_\Sigma$ is just an injective Lie algebroid over $\Sigma$
with the anchor $\bar R=R|_\Sigma$. The constructions of Sec.2
correspond to the special case where both $E_1$ and $E_2$ are
trivial vector bundles.

In  view of an apparent similarity to the definition of the Lie
algebroid, it is natural to call the quadruple $(E_1,E_2,R, T)$ a
\textit{weak Lie algebroid}, as the integrability of the (anchor)
distribution $R$ takes place only on the (constraint) surface
$\Sigma=\{x\in M| T(x)=0\}$.

Now let us describe the BRST-imbedding of a gauge system involving
a weak Lie algebroid $(E_1,E_2,R, T)$ and  a  projectible vector
field $V\in \Lambda^1(M)^{\mathrm{pr}}$ respecting a weak Poisson
structure $P\in \Lambda^2(M)^{\mathrm{pr}}$. In this case, the
supermanifold of fields $\mathcal{M}$ is chosen to be the direct
sum $\Pi E_1\oplus \Pi E_2$ of odd vector bundles over $M$. Let
$\eta_a$ and $c^\alpha$ denote linear coordinates in fibers of
$\Pi E_1$ and $\Pi E_2$ over a trivializing chart $U\subset M$
with local coordinates $x^i$.  The whole field-antifield
supermanifold $\mathcal{N}$ is associated with the total space of
the vector bundle (cf. (\ref{dpr}))
\begin{equation}\label{dsum}
E^\ast_1\oplus E_2^\ast\oplus \Pi E_1\oplus \Pi E_2\oplus \Pi
T^\ast M\,.
\end{equation}

Choosing a linear connection $\nabla=\nabla_1\oplus\nabla_2$ on
$E_1\oplus E_2$, one can endow $\mathcal{N}$ with an exact
antisymplectic structure $\Omega=d\Theta$, where
\begin{equation}\label{}
    \Theta = (\sx_idx^i+\se^a\nabla_1 \eta_a + \sc_\alpha \nabla_2c^\alpha)\,,
\end{equation}
$$
\nabla_1\eta_a=d\eta_a+dx^i\Gamma_{i a}^b(x)\eta_b\,,\qquad
\nabla_2c^\alpha=dc^\alpha+dx^i\Gamma_{i\beta}^\alpha(x)c^\beta\,.
$$
The corresponding antibrackets of fields $\phi^A=(x^i, c^\alpha,
\eta_a)$ and antifields $\phi^\ast_A=(\sx_i,\sc_\alpha, \se^a)$
read
\begin{equation}\label{brcov}
\begin{array}{ll}
 (\sc_\alpha, c^\beta)=\delta_\alpha^\beta \,,\quad&  (\sx_i,c^\alpha)=\Gamma_{i\beta}^\alpha c^\beta
 \,,\qquad
 (\sx_i,\sc_\alpha)=-\Gamma_{i\alpha}^\beta \sc_\beta\,, \\[3mm]
 (\se^a,\eta_b)=\delta_b^a\,,\quad&(\sx_i,\eta_a)=\Gamma_{ib}^a\eta_b\,\,,\,\qquad
 (\sx_i,\se^a)= -\Gamma_{i b}^a \se^b\,,
  \\[3mm]
 (\sx_i, x^j)=\delta_i^j\,, \quad& (\sx_i,\sx_j)=R_{ij\beta}^\alpha
\sc_\alpha
  c^\beta+R_{ija}^b\se^a\eta_b\,,
\end{array}
\end{equation}
and the other brackets vanish. Here $R(x)_{ij\beta}^\alpha$ and
$R(x)_{ija}^b$ are the curvatures of $\nabla_1$ and $\nabla_2$.

As with the case of trivial vector bundles $E_1$ and $E_2$,
considered in Sec. 3, all the ingredients of the gauge system are
combined into the pair of functions $S$ and $H$ subject to the
master equations (\ref{me}). The covariance of the antibrackets
(\ref{brcov}) under the bundle automorphisms suggests  the
coefficients of the expansions (\ref{S}, \ref{H}) for $S$ and $H$
to transform homogeneously i.e. as the tensors associated with the
vector bundle (\ref{dsum}).

Since the antibrackets (\ref{brcov}) have not the canonical form,
we cannot apply the formality theorem directly to perform the
deformation quantization of the gauge system as it has been done
in Sec.4. The generalization of the formality theorem to arbitrary
antisymplectic manifolds (which are not odd cotangent bundles with
the canonical antibrackets) still remains an interesting open
question. For lack of a rigorous mathematical treatment we will
use a less rigor but more descriptive approach of topological
sigma-models of Ref. \cite{CF}, where it was originally used to
elucidate the Kontsevich quantization formula for Poisson
manifolds. Below, we briefly explain how this approach works in a
more general situation of the weak Poisson structure.

Following  \cite{bm},  \cite{CF}, \cite{AKSZ},   we consider a
topological sigma-model having $\mathcal{N}$ as target space, and
whose world sheet is given by the supermanifold $\Pi TU$, where
$U$ is a closed two-dimensional disk.  Let  $u^\mu$, $\mu=1,2$,
denote local coordinates on $U$ and $\theta^\mu$ denote odd
coordinates on the fibers of $\Pi TU$. The natural volume element
on $\Pi TU$ is given by $d\mu=d^2ud^2\theta$. We choose the odd
coordinates $\theta^\mu$ to have ghost number $1$.

Let $\Phi^I=(\phi^A,\phi^\ast_B)$ be a collective notation for all
the fields and antifields. Each field configuration
$\Phi^I(u,\theta)$ defines an imbedding of $\Pi TU$ into
$\mathcal{N}$. Expanding superfield $\Phi^I$ in $\theta$'s one
gets a collection of ordinary fields on $U$,
\begin{equation}\label{cfil}
\Phi^I(u,\theta)=\Phi_0^I(u)+\theta^\mu
\Phi^I_\mu(u)+\theta^\mu\theta^\nu\Phi^I_{\mu\nu}(u)\,,
\end{equation}
The parities and ghost numbers of the component fields $\Phi^I_0,
\Phi^I_\mu, \Phi^I_{\mu\nu}$ are uniquely determined by those of
$\Phi^I$ and $\theta^\mu$. Each superfield $\Phi^I$ is naturally
identified with an inhomogeneous differential form on $U$; in
doing so, the role of exterior differential is played by the
supercovariant derivative
\begin{equation}\label{}
 \mathrm{D}=\theta^\mu\frac{\partial}{\partial u^\mu}\,,\quad\Rightarrow\quad \mathrm{D}^2=0\,.
\end{equation}

With any solution $S(\Phi)$ of the master equation $(S,S)=0$ one
can associate the sigma-model \cite{AKSZ}, \cite{bm}, whose master
action reads
\begin{equation}\label{a}
    \mathcal{A}[\Phi]=\int_{\Pi TU} d\mu \left(\Theta
    _I(\Phi)\mathrm{D}\Phi^I-S(\Phi)\right)\,,\qquad
    \gh(\mathcal{A})=0\,.
\end{equation}
The antibrackets (\ref{brcov}) on the target space $\mathcal{N}$
give rise to antibrackets on the configuration space of
superfields:
\begin{equation}\label{brc}
(F,G)'= \int_{\Pi TU} d\mu \left(
\frac{F\overleftarrow{\delta}}{\delta\Phi^I(u,\theta)}
\Omega^{IJ}\frac{\overrightarrow{\delta}G}{\delta
\Phi^J(u,\theta)}\right)\,,\qquad\Omega_{IJ}\Omega^{JK}=\delta_I^K\,,
\end{equation}
for any functionals $F[\Phi]$ and $G[\Phi]$.  By construction, the
action (\ref{a}) obeys the classical master equation
\begin{equation}\label{}
    (\mathcal{A},\mathcal{A})'=2\int_{\Pi TU} d\mu
    \mathrm{D}[\Theta(\Phi)_I\mathrm{D}\Phi^I+S(\Phi)] + \int_{\Pi TU} d\mu (S,S)(\Phi)=0\,.
\end{equation}
The second term vanishes due to the classical master equation
$(S,S)=0$, while vanishing of the first term is due to appropriate
boundary conditions on the component fields (\ref{cfil}). In
particular, it is assumed that
\begin{equation}\label{bcon}
\phi^\ast_A|_{\partial U}=0\,.
\end{equation}

The BV quantization of the model (\ref{a}) requires an integration
measure (or density) to be introduced on the configuration space
of fields $\Phi^I(u,\theta)$. As well as the antibrackets, the
measure can be obtained from a normal, nowhere vanishing density
on the target space $\mathcal{N}$. Recall that a density $\rho$ is
said to be \textit{normal} if there is an atlas of Darboux's
coordinates (\ref{abr}) on $\mathcal{N}$ in which $\rho=1$. Any
normal density gives rise to the nilpotent odd Laplace operator
$\Delta: C^{\infty}(\mathcal{N})\rightarrow
C^{\infty}(\mathcal{N})$ :
\begin{equation}\label{df}
\Delta f=\mathrm{div}_\rho X_f\,,\qquad \Delta^2=0\,,
\end{equation}
$X_f=(f,\,\cdot\,)$ being the Hamiltonian vector field
corresponding to $f$.

Given a normal density $\rho$ on $\mathcal{N}$, the natural
integration measure on the configuration space of fields can be
chosen as
\begin{equation}\label{}
\mathcal{D}\Phi=\rho'[\Phi]\prod_{z\in \Pi TU}\delta
\Phi^1(z)\cdots \delta\Phi^N(z)\,,\qquad \rho'[\Phi]=\prod_{z\in
\Pi TU}\rho(\Phi(z))\,,
\end{equation}
where $N=\dim \mathcal{N}$. The functional counterpart of the odd
Laplacian (\ref{df}) is then given by
\begin{equation}\label{lap}
\Delta' =\int_{\Pi TU}d\mu \,\rho'^{-1}\frac{\delta}{\delta
\Phi^I(u,\theta)}\rho'\Omega^{IJ}\frac{\delta}{\delta
\Phi^J(u,\theta)}\,,
\end{equation}
Although, in the field-theoretical context, the odd Laplacian is
known to be an ill-defined operator, we claim that after an
appropriate regularization
\begin{equation}\label{}
    \Delta' \mathcal{A}=0\,.
\end{equation}
The reason is as follows. Since $\Delta'$ is a local operator, one
can compute its action passing to suitable Darboux's coordinates
on $\mathcal{N}$ in  which $\rho' = 1$. In these coordinates,
$\Delta'$ becomes a homogeneous second order operator.
Consequently,
\begin{equation}\label{d0}
    \Delta' \mathcal{A} = \delta(0)(\mathrm{something})\,,
\end{equation}
where the $\delta$-function on $\Pi TU$ is defined as $\delta (z)=
\theta^1\theta^2\delta(u^1)\delta(u^2)$. So, the  right hand side
of (\ref{d0}) vanishes due to the $\theta$-part of the
$\delta$-function.

As a consequence the classical master action $\mathcal{A}$ obeys
also the quantum master equation
\begin{equation}\label{qme}
    \frac12(\mathcal{A},\mathcal{A})'=\hbar\Delta'\mathcal{A}\,.
\end{equation}

By definition, the BRST observables are functionals
$\mathcal{O}[\Phi]$ annihilated by the quantum BRST operator
$\Omega$:
\begin{equation}\label{brsto}
    \Omega \mathcal{O}\equiv -i\hbar\Delta'
    \mathcal{O}+(\mathcal{A},\mathcal{O})'=0\,.
\end{equation}
Using Eq.(\ref{qme}) and the fact that $\Delta'$ is a derivation
of the antibrackets (\ref{brc}) of degree 1, one can see that
$\Omega^2=0$. A particular class of BRST observables is provided
by the $D$-closed functions from $C^\infty(\mathcal{M})$ of ghost
number zero. (Recall that the cohomology classes of these cocycles
are precisely identified  with the classical observables of our
gauge system.) Let us set $\mathcal{O}_f(u)= f(\phi(u,0))$ for
$u\in
\partial U$. From the definition (\ref{Dcl}) of a $D$-cocycle and the
boundary condition (\ref{bcon}) for antifields it then follows
that
\begin{equation}\label{}
(\mathcal{A}, \mathcal{O}_f(u))'= (S, f)(\phi(u,0)) = (Q,
f)(\phi(u,0))=0\,.
\end{equation}
Furthermore, passing to suitable  Darboux's coordinates  one can
also see that $
    \Delta' \mathcal{O}_f(u)\sim \delta(0)$ and hence, after an
    appropriate regularization, $\Delta'\mathcal{O}_f(u)=0$.
    So, any classical observable gives rise to the BRST observable.

By analogy with \cite{CF}, we define the weakly associative
$\ast$-product of two classical observables $f, g\in
C^{\infty}(\mathcal{M}) $ by the following path integral:
\begin{equation}\label{stpr}
(f\ast g)(x)=\int_{\mathcal{L}} \mathcal{D}\Phi \mathcal{O}_f(
u)\mathcal{O}_g(v)\delta[\phi(w)-x]e^{\frac i\hbar
\mathcal{A}[\Phi]}
\end{equation}
Here $u,v,w$ are three distinct points on the boundary circle
$\partial U$, and the integral is taken over a suitable Lagrangian
submanifold  $\mathcal{L}\subset\mathcal{N}$. Being a BRST
observable, the integrand (\ref{stpr}) is annihilated by the odd
Laplacian (\ref{lap}) and  the weak associativity of the
$\ast$-product follows formally from the usual factorization
arguments (see e.g. \cite{CF}).

Non-rigor though the path-integral arguments are, they will
hopefully be found useful as a  practical tool and as a starting
point for constructing a more rigor deformation quantization of
generic (not necessarily Hamiltonian) gauge systems whose BRST
embedding is proposed in this paper.

\section*{Acknowledgments}
We are indebted to V.A. Dolgushev for various informal discussions
of the formality theorem. We are thankful to I.V. Tyutin for
useful remarks on the work and to M.A. Grigoriev for the reference
\cite{BGST} which we have overlooked. This work benefited from the
following research grants: Russian Ministry of Education grant E
02-3.1-250, and the grant for Support of Russian Scientific
Schools 1743.2003.2. AAS appreciates the financial support from
the Dynasty Foundation and the International Center for
Fundamental Physics in Moscow, and from the RFBR grant
03-02-17657. SLL is partially supported by the RFBR grant
02-01-00930.

\end{document}